\documentclass{PoS}

\title{Inclusive search for boosted Higgs bosons using H$ \rightarrow \mathrm{b\overline{b}}$ decays with the CMS experiment}

\ShortTitle{ggF H$\rightarrow\mathrm{b\overline{b}}$}

\author{\speaker{Caterina Vernieri}\thanks{on behalf of the CMS Collaboration}\\
       Fermi National Accelerator Laboratory, Batavia, IL, USA,\\
        E-mail: \email{caterina.vernieri@cern.ch}}

\abstract{ We present the first search for the standard model Higgs boson (H) produced with large transverse momentum ($\mathrm{p_{T}}$) via gluon fusion and decaying to a bottom quark-antiquark pair ($\mathrm{b\overline{b}}$). The search is performed using a data set of pp collisions at $\sqrt{s}=13$~TeV collected with the CMS experiment at the LHC, corresponding to an integrated luminosity of 35.9~fb$^{\mathrm{-1}}$. A highly Lorentz-boosted Higgs boson decaying to $\mathrm{b\overline{b}}$ is reconstructed as a single, large radius jet and is identified using jet substructure and dedicated b tagging techniques. The method is validated with the first observation of  the Z$\rightarrow\mathrm{b\overline{b}}$ process in the single-jet topology, with a local significance of 5.1 standard deviations (5.8 expected). }

\FullConference{The European Physical Society Conference on High Energy Physics\\
		5-12 July, 2017\\
		Venice}

\begin{document}

\section{Motivations}

At a mass ($\mH$) of 125 GeV the standard model (SM) Higgs boson decay mode into a bottom quark-antiquark pair (\bbbar) dominates the total width~(${\sim}58\%$)~\cite{LHCHiggsCrossSectionWorkingGroup:2011ti}. While the decay of the Higgs boson to vector bosons has been observed in different channels (ZZ, $\gamma\gamma$,WW)~\cite{ZZ,Hgg,WW}, the direct couplings of the H to down-type quarks, remains to be firmly established~\cite{Nature-14,Aaboud:2017xsd,Tevatron,CMSevidence}. 

The current measurements constrain indirectly the couplings to the up-type top quark, since the dominant Higgs boson production mechanism is gluon fusion induced by top-quark loop. The measurement of the H$(\bbbar)$ decay represents a direct test of whether the observed boson interacts as expected with the quark sector and provides the unique final test of the direct coupling of the Higgs boson to down-type quarks, an essential aspect of the nature of the newly discovered boson.  To date, the most precise constraints on the couplings to down-type quarks are provided by the CMS and ATLAS experiments, that have recently announced the first evidence for the Higgs boson decay into b quarks and for its production in association with a vector boson~\cite{Aaboud:2017xsd,CMSevidence}. No one has yet to consider the gluon fusion (ggF) production mode, despite being the dominant production mechanism at the LHC (87\%). The search for ggF H(\bbbar) historically has been deemed impossible~\cite{ZbbTevatron,ZbbAtlas} because of the overwhelming irreducible background from QCD production of b quark, which is roughly 7 order of magnitude larger than the signal. 
We present the first inclusive search for H$\to\bbbar$ based on a data set of pp collisions at $\sqrt{s} = 13$~TeV collected with the CMS detector~\cite{CMSdet} at the LHC in 2016 and corresponding to an integrated luminosity of 35.9~\fbinv~\cite{HIG-17-010}.
The main experimental difficulties for this search originate from the large cross section for background multijet events and the restrictive
trigger requirements needed to reduce the data recording rate.
Therefore, we require events to have a high-$\pt$ H, above 450 GeV. 
Being this search performed at very high $\pt$ it could also potentially be sensitive to new physics variation to the couplings that would enhance or reduce the Higgs boson production cross section~\cite{grazzini,maltoni}, if the scale of new physics is larger than the electroweak scale.

\section{ggF H modeling at high \pt}
Computing the differential cross section in H $\pt$ for ggF H $\pt > 450 $~GeV poses a number of challenges. 
At low H $\pt$, the dominant contributions come from the application of higher order corrections which are large for loop-induced processes. The dominant correction at values of the H $\pt$ greater than approximately twice the mass of the top quark ($m_{t}$) originates from the resolved top quark loop (finite top mass correction)~\cite{Baur:1989cm}. The resolved top quark loop induces a deficit in the production of Higgs bosons at high $\pt$ relative to the case where the loop is unresolved, known as the effective field theory (EFT) or $m_{t}\rightarrow\infty$ approximation.

\begin{figure}[hbtp]\begin{center}
    \includegraphics[width=0.46\textwidth]{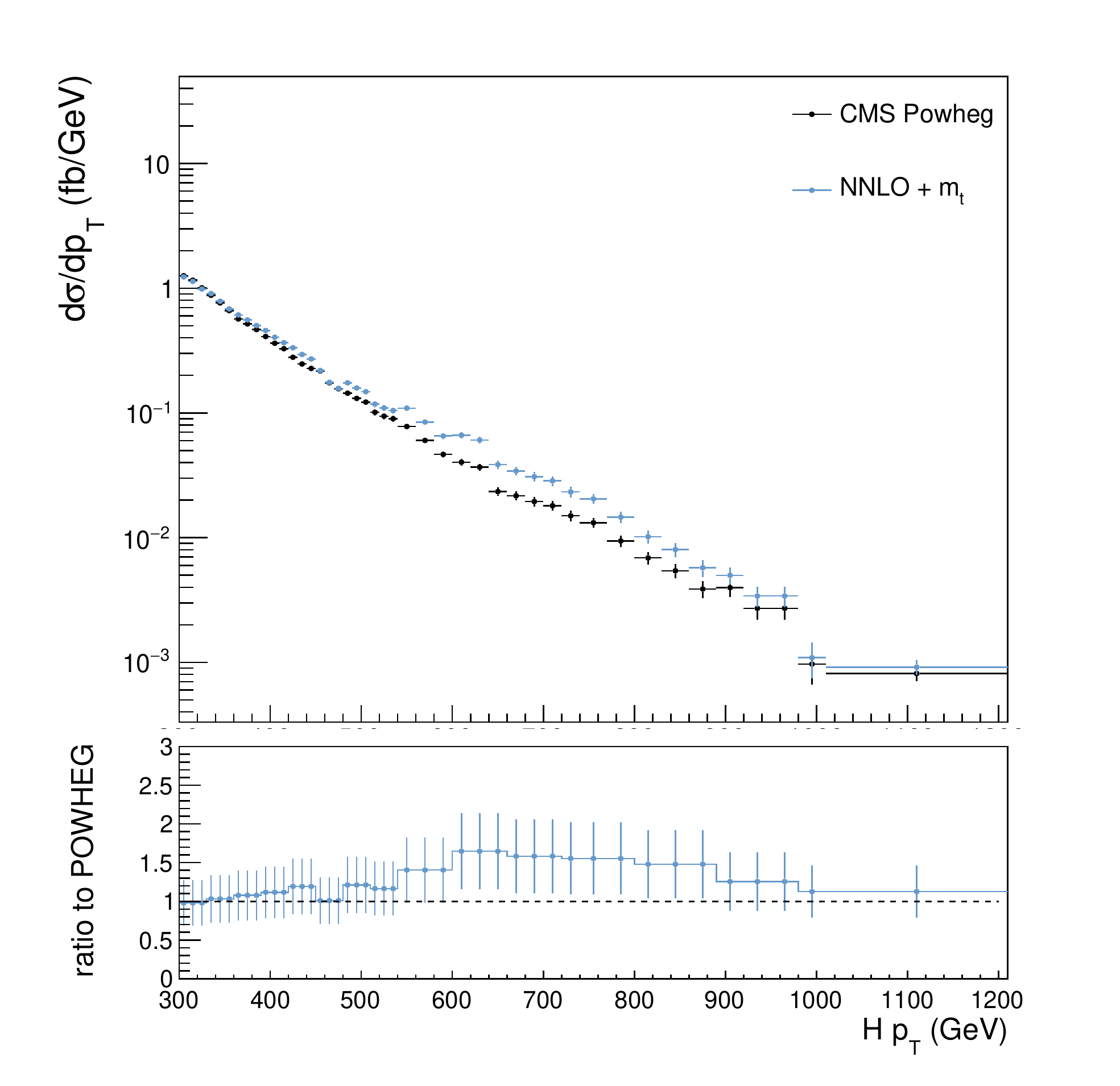}
    \caption{Generator level H $\pt$ distribution for the gluon fusion production mode. The CMS default {\sc Powheg} sample and the corrected spectrum to account for both higher order and finite top mass effects are compared~\cite{HIG-17-010}.}
 \label{fig:Higgspt}
 \end{center}
 \end{figure}

In the interest of comparing with other CMS results, the {\sc POWHEG} generator with H matrix elements up to 1 jet is used and tuned with the \emph{h-fact} parameter set to $104.13$~GeV. The resulting tuned H generation is normalized to the inclusive next-to-next-to-next-to-leading order (N$^{3}$LO) accuracy.
In addition, an alternative approach is considered to get the highest order possible differential H $\pt$ spectrum~\cite{Boughezal:2013uia,Boughezal:2016wmq,Chen:2016vqn}, while preserving the finite top mass correction~\cite{Neumann:2016dny}.  To account for both the effects of higher order corrections and the resolved top loop, a multi-correction approach is adopted~\cite{Buschmann:2014sia,Frederix:2016cnl}, that can be summarized as:
\begin{equation}
\label{eq:higgspt}
\small{\mathrm{ggF~H}\mathrm{(NNLO+m_{t})}}   = \small{\mathrm{Powheg}(1~\mathrm{jet}~m_{t}) \times \frac{\mathrm{MG~LO}~0-2~\mathrm{jet}~m_{t}}{\mathrm{Powheg}(1~\mathrm{jet}~m_{t})} \times}  \small{\frac{\mathrm{NLO}~1~\mathrm{jet}~m_{t}}{\mathrm{LO}~1~\mathrm{jet}~m_{t}} \times \frac{\mathrm{NNLO}~1~\mathrm{jet}~m_{t}\rightarrow\infty}{\mathrm{NLO}~1~\mathrm{jet}~m_{t}\rightarrow\infty}}
\end{equation}
Samples are generated at LO for the 0, 1, and 2 jet H production with {\sc MADGRAPH} (MG) using the loop$_\mathrm{SM}$ model~\cite{Hirschi:2015iia} showered with CKKW-L scheme~\cite{Catani:2001cc,Lonnblad:2001iq,Czakon:2013goa}. This spectrum is then corrected by the approximate NLO to LO ratio, obtained by expanding in powers of $1/m_{t}^2$ and it is found to be $2.0\pm0.5$ and roughly constant as a function of $\pt$. 
The effective NNLO to NLO ratio~\cite{Neumann:2016dny} in the infinite top quark mass approximation is found to be $1.25\pm0.15$ and is also roughly constant across $\pt$~\cite{Chen:2014gva,Boughezal:2013uia,Boughezal:2015dra}.
For H $\pt > 450 $~GeV, the correction to the default {\sc POWHEG} is found to be $1.27\pm0.38$, resulting in a cross section of $31.7\pm9.5$~fb for ggF H$\to\bbbar$. The Higgs boson generator-level $\pt$ distribution is shown in Fig.~\ref{fig:Higgspt}, to compare the {\sc POWHEG} prediction and the one derived to account of higher order corrections and the finite top mass loop.
An  uncertainty of 30\% to the overall correction is estimated from the comparison of different predictions obtained by using: (i) a merging scale of 100 instead of 20~GeV, (ii) the inclusive two-jet ggF process generation, (iii) the {\sc MadGraph5\_amc@nlo} effective field theory approximation~\cite{Neumann:2016dny,deFlorian:2016spz} normalized to the inclusive N$^{3}$LO cross section.


%
%
%
%
%
%

\section{Higgs boson tagging}

The angular separation of the two b from H decay can be approximated by 2$\mH/\pt$. Thus, at high $\pt$ the products of hadronization of two b quarks are merged in one single large cone size (fat) jet. 
Higgs boson candidates decaying to $\bbbar$ are reconstructed as jet (H jet) using the anti-k$_{\rm T}$ algorithm~\cite{Cacciari:2008gp} with a distance parameter of 0.8 (AK8 jets), and specific techniques are used to mitigate effects from multiple p-p interactions per collision (pileup). The pileup per particle identification (PUPPI) algorithm~\cite{Bertolini:2014bba} assigns a weight to each particle prior to jet clustering based on the likelihood of the particle originating from the hard interaction. Mainly three properties are then exploited to isolate H jet candidate from background of q/g jets: the jet mass, b tagging and the composite nature of the jet using substructure.


The jet mass should be compatible with the Higgs boson mass hypothesis and provide good discrimination from quark- and gluon-initiated jets. Soft-drop grooming is applied to remove soft and wide-angle radiation~\cite{Dasgupta:2013ihk,Larkoski:2014wba}. Grooming tends to push the jet mass scale of the background to lower values while preserving the hard scale of the heavy resonance. The soft drop jet mass ($\mSD$) peaks at the H mass for signal events and reduces the masses of jets from background. 
The $N_2^{1}$ variable~\cite{Moult:2016cvt}, which is based on a ratio of 2-point and 3-point generalized energy correlation functions (ECFs)~\cite{Larkoski:2013eya}, is exploited to determine how consistent a jet is with having a two-prong substructure. However, any selection on $N_2^{1}$ shapes the jet mass distributions differently depending on the $\pt$ of the jet. Therefore a mass-decorrelated procedure is applied~\cite{Dolen:2016kst} to achieve a constant QCD background efficiency of 26\% across the entire mass and $\pt$ range considered in this search. The chosen percentile maximizes the sensitivity to the H($\bbbar$) signal.

In order to select events in which the H jet is most likely to contain two b quarks, we use the double-b tagger algorithm~\cite{CMS-PAS-BTV-15-002}.
Several observables that characterize the distinct properties of b hadrons and their flight directions in relation to the jet substructure are used as input variables to this multivariate algorithm, to distinguish between H jets and QCD jets. In this search, an H jet is considered double-b tagged if its double-b tag discriminator value is above a threshold corresponding to a 1\% misidentification rate for QCD jets and a 33\% efficiency for H$(\bbbar)$ jets.  By design the mistag rate is approximately flat across the $\pt$ range, and it is a critical point for this search. 

\section{Event Selection}
We exploit the H tagging tools to perform the first search for ggF H($\bbbar$). We look for a single high-$\pt$ H jet, recoiling against some other object, like a narrow jet radiated in the initial state, although no assumption or requirement on such additional object is actually made in the analysis.
Combinations of several online selections are used, all requiring the total hadronic transverse
energy in the event (H$_{\rm T}$) or jet $\pt$ to be above a given threshold.
The online selection is fully efficient at selecting events offline
with at least one AK8 jet with $\pt > 450$~GeV and $|{\eta}| < 2.5$.
The leading (in $\pt$) jet in the event is assumed to be the Higgs boson candidate, the H jet, and the substructure and b tagging requirements are applied.
Events with (without) a double-b tagged H jet define the passing (failing)
region.
In the passing region, the gluon fusion process dominates,
although other Higgs boson production
mechanisms contribute: VBF (12\%), VH (8\%), $\ttbar$H (5\%). They are all taken into account when extracting the Higgs boson yield.

%
\section{Background modeling}

The contribution of $\ttbar$ production to the total SM
background is estimated to be less than 3\%. It is obtained from simulation
corrected with scale factors derived from a $\ttbar$-enriched control
sample in which an isolated muon is required.

The contributions of the W and Z$+$jets processes is about 5\% and it is modeled from simulation. 
During the signal extraction we measure the Z$+$jets normalization while the W$+$jets normalization is allowed to vary within its systematic uncertainties.


The main background in the passing region,
QCD multijet production, has a nontrivial jet mass shape that is difficult to model parametrically and dependent on jet $\pt$
so we constrain it using the signal-depleted failing region.
Since the double-b tagger discriminator and the jet mass are largely uncorrelated, the passing and failing
regions have similar QCD jet mass distributions, and their ratio,
the ``pass-fail ratio'' $R_{\mathrm{p}/\mathrm{f}}$, is expected to
be nearly constant as a function of jet mass and $\pt$.
To account for the residual difference between the
shapes of passing and failing events, $R_{\mathrm{p}/\mathrm{f}}$ is
parametrized as a polynomial in $\rho$ and $\pt$, which is determined from a simultaneous fit to the data in passing and failing regions across the whole jet mass range.
\section{Results}
\begin{figure}
\centering
    \includegraphics[width=0.45\textwidth]{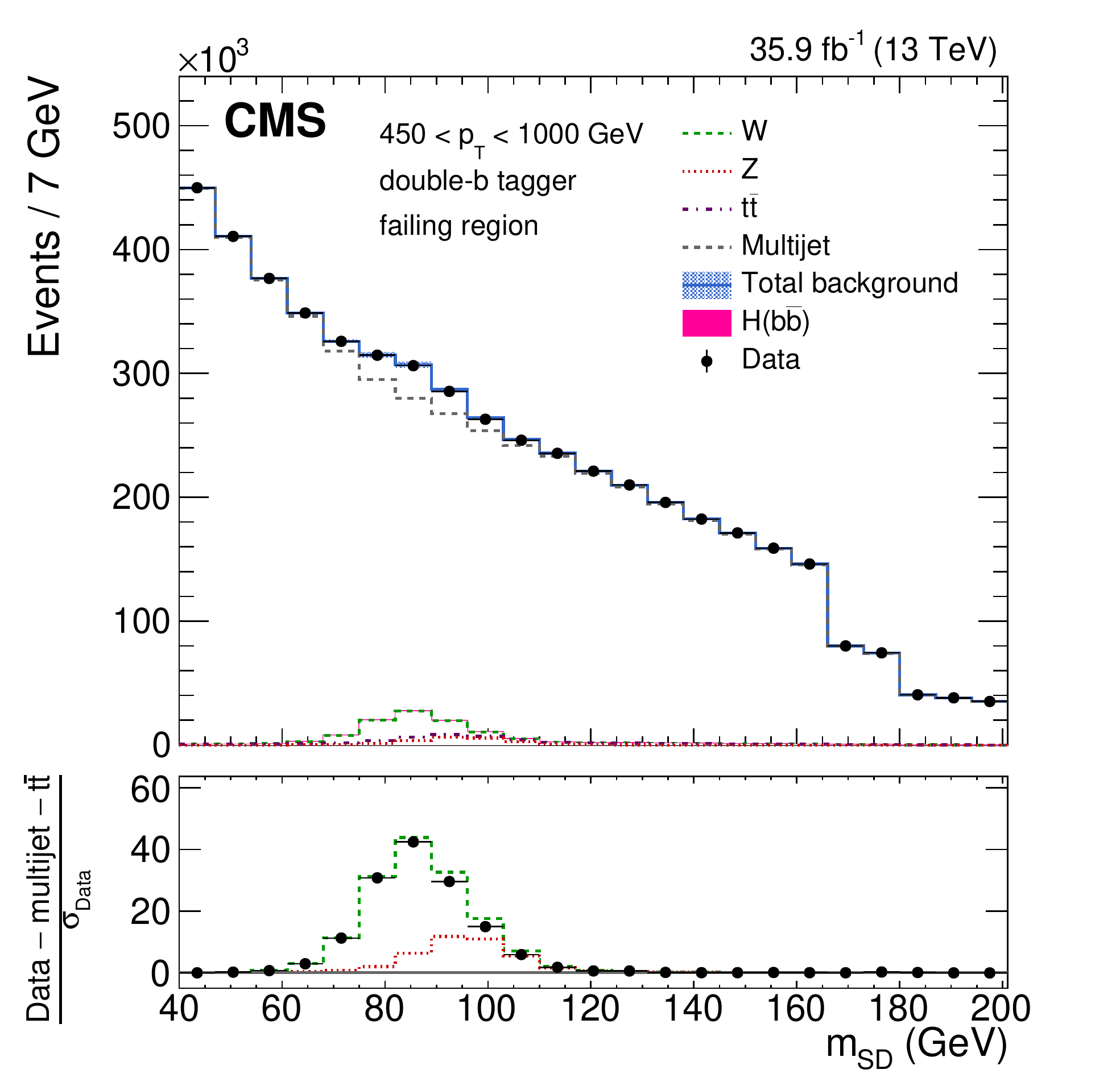}
    \includegraphics[width=0.45\textwidth]{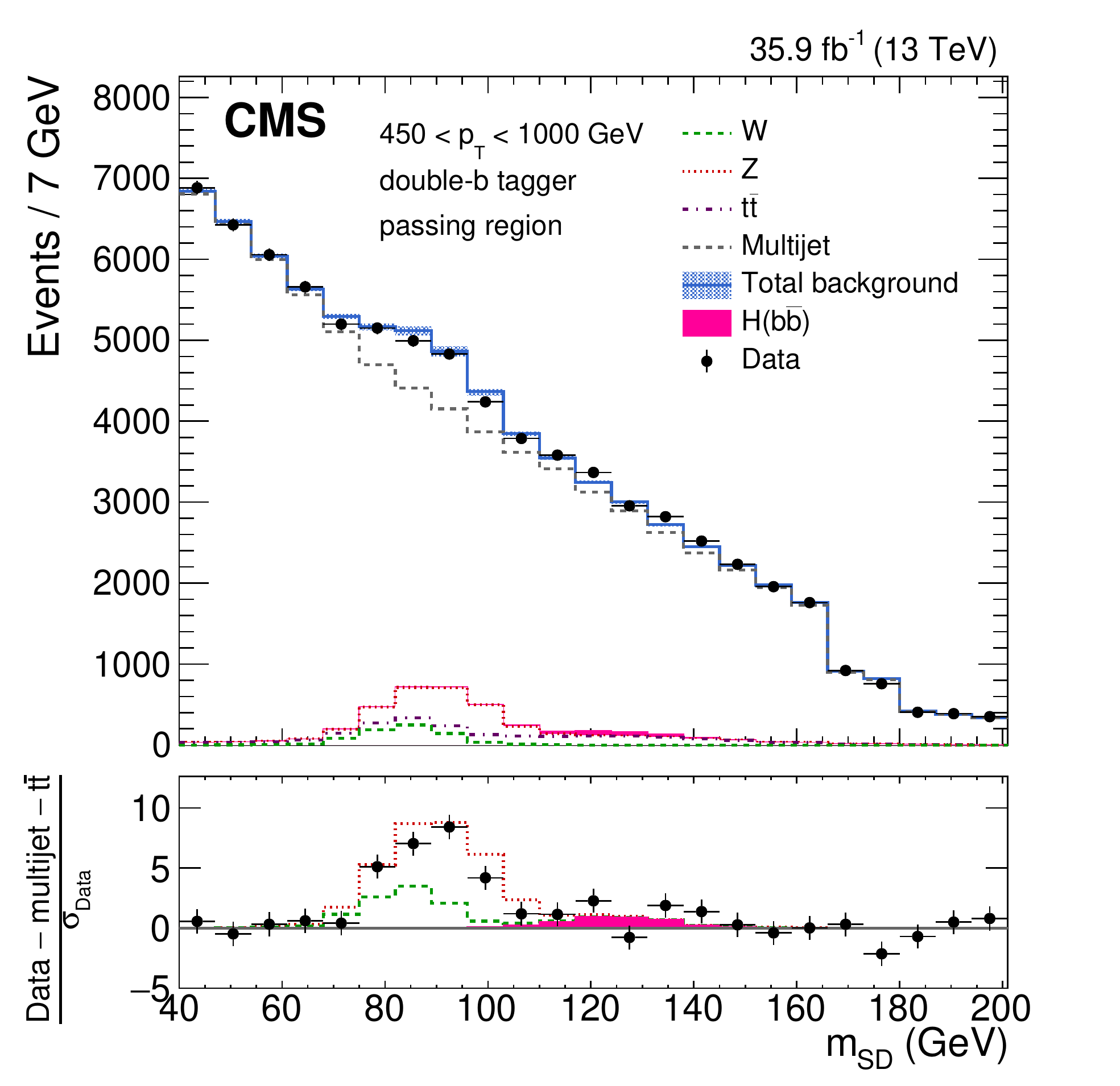}
 \caption{The $\mSD$ distributions in data for the
      failing (left) and passing (right) regions.
      In the bottom panel, the ratio of the data to its statistical uncertainty, after subtracting the non-resonant backgrounds, is shown~\cite{HIG-17-010}.}
 \label{fig:results}
 \end{figure}

A binned maximum likelihood fit to the observed $\mSD$ distributions in the range 40 to 201~GeV is performed
using the sum of the H$(\bbbar)$, W, Z, $\ttbar$, and QCD multijet contributions.
The fit is done simultaneously in the passing and failing regions. The production cross sections relative to the SM cross sections (signal strengths) for the Higgs and the Z bosons, $\mu_{\rm H}$ and $\mu_{\rm Z}$, respectively, are extracted from the fit.
Figure~\ref{fig:results} shows the $\mSD$
distributions in data for the passing and failing regions with
measured SM background and H$(\bbbar)$ contributions. Contributions
from W and Z boson production are clearly visible in the data.
Table ~\ref{tab:ObservedSig} summarizes the measured signal strengths and significances for the Higgs and Z boson processes. In particular, they are also reported for the case in which no corrections to the Higgs boson $\pt$ spectrum are applied.
%
%
\begin{table}[htbp]
\centering
 \caption{Fitted signal strength, expected and observed significance of the Higgs and Z boson signal. }
\begin{tabular}{ llll }
\hline
 & H & H no \pt~ corr. & Z \\\hline
Observed signal strength & $\muVal_{-\muErrLo}^{+\muErrHi}$ & $\muNoVal_{-\muNoErrLo}^{+\muNoErrHi}$  & $\muZVal_{-\muZErrLo}^{+\muZErrHi}$\\\hline
Expected significance & $\muExpSig\sigma$ & $\muNoExpSig\sigma$ &  $\muZExpSig\sigma$ \\\hline
Observed significance & $\muObsSig\sigma$ & $\muNoObsSig\sigma$ &  $\muZObsSig\sigma$\\\hline
\end{tabular}
\label{tab:ObservedSig}
\end{table}
\section{Conclusions}

In summary, an inclusive search for the standard model Higgs boson with $\pt > 450 $~GeV decaying to bottom quark-antiquark pairs and reconstructed as a single, large-radius jet is presented.
The Z$+$jets process is observed for the
first time in the single-jet topology with a significance of
$\muZObsSig\sigma$. The Higgs boson production is measured
with an observed (expected) significance of $\muObsSig\sigma$
($\muExpSig\sigma$) when including Higgs boson $\pt$ spectrum corrections accounting for higher-order and finite top quark mass effects. The measured cross section times branching fraction for the gluon fusion H$(\bbbar)$ production for $\pt>450$~GeV and $|{\eta}|<2.5$ is
$\xsecVal\pm\xsecStatErrLo{\rm(stat)}_{-\xsecSystErrLo}^{+\xsecSystErrHi}{\rm(syst)}$~fb,
which is consistent with the SM prediction within uncertainties.
This search looks at previously unexplored regions of phase space and opens a new strategy to search for H($\bbbar$) and probe BSM contributions to the Higgs boson production cross section at very high $\pt$.

%



\begin{thebibliography}{99}
 \bibitem{LHCHiggsCrossSectionWorkingGroup:2011ti}{LHC Higgs Cross Section Working Group CERN-2011-002, arXiv:1101.0593 (2011)}

 \bibitem{ZZ}{CMS Collaboration, Phys. Rev. D 89 (2014) 092007}

\bibitem{Hgg}{CMS Collaboration, EPJC 74 (2014) 3076}
  
\bibitem{WW}{CMS Collaboration,  JHEP 1401 (2014) 096}
                        
 \bibitem{Nature-14}{CMS Collaboration,  Nature Phys. 10 (2014) 3005}
 \bibitem{Aaboud:2017xsd}{ATLAS Collaboration,  arXiv:1708.03299 (2017) submitted to JHEP}  
  \bibitem{Tevatron}{CDF and D0 Collaborations, Phys. Rev. D {88} (2013) 052014}

 \bibitem{CMSevidence}{CMS Collaboration, arXiv:1709.07497 (2017) submitted to PLB}
 \bibitem{ZbbTevatron}{CDF Collaboration, CDF-PUB-11228 (2017) }
 \bibitem{ZbbAtlas}{ATLAS Collaboration, Phys. Lett. B 738 (2014)}
\bibitem{CMSdet}{CMS Collaboration, JINST 3 (2008) S08004}
 \bibitem{HIG-17-010}{CMS Collaboration, arXiv:1709.05543 (2017) submitted to PRL}
 \bibitem{grazzini}{M.~Grazzini, A.~Ilnicka, M.~Spira and M.~Wiesemann,  JHEP 1703 (2017) 115}
 \bibitem{maltoni}{F.~Maltoni, E.~Vryonidoua and C.~Zhangb, JHEP 10 (2016) 123}
 \bibitem{Baur:1989cm}{U.~J.~Baur and E. W. N.~Glover, Nucl. Phys. B 339 (1990) 38}
 \bibitem{Boughezal:2013uia}{R.~Boughezal, F.~Caola, K.~Melnikov, F.~Petriello and M.~Schulze, JHEP 06 (2013) 072}
 \bibitem{Boughezal:2016wmq}{R.~Boughezal, J.~M.~Campbell, R.~K.~Ellis, C.~Focke, W.~Giele, X.~Liu, F.~Petriello, and C.~Williams, arXiv:1605.08011 (2016)}
 \bibitem{Chen:2016vqn}{X.~Chen, T.~Gehrmann, N.~Glover, and M.~Jaquier, arXiv:1604.04085 (2016)}
  \bibitem{Neumann:2016dny}{T.~Neumann, and C.~Williams, Phys. Rev. D 95 (2017) 014004}
  \bibitem{deFlorian:2016spz}{D.~de Florian, C.~Grojean, F.~Maltoni, C.~Mariotti, A.~Nikitenko, M.~Pieri, P.~Savard, M.~Schumacher and R.~Tanaka, arXiv:1610.07922 (2016)}
 \bibitem{Buschmann:2014sia}{M.~Buschmann,  D.~Goncalves, S.~Kuttimalai, M.~Schonherr, F.~Krauss, and T.~Plehn, JHEP 02 (2015) 038}
 \bibitem{Frederix:2016cnl}{R.~Frederix,  S.~Frixione, E.~Vryonidou and M.~Wiesemann, JHEP 08 (2016) 006}
 \bibitem{Hirschi:2015iia}{V.~Hirschi and O.~Mattelaer, JHEP 10 (2015) 146}
 \bibitem{Catani:2001cc}{S.~Catani, F.~Krauss, R.~Kuhn, B.~R.~Webber, JHEP 11 (2001) 063}
 \bibitem{Lonnblad:2001iq}{L.~Lonnblad, JHEP 05 (2002) 046}
 \bibitem{Czakon:2013goa}{M.~Czakon, P.~Fiedler and A.~Mitov, Phys. Rev. Lett. 110 (2013) 252004}
 \bibitem{Chen:2014gva}{X.~Chen, T.~Gehrmann, E.~W.~N.~Glover, and M.~Jaquier, Phys. Lett. B 740 (2015)}
 \bibitem{Boughezal:2015dra}{R.~Boughezal, F.~Caola, K.~Melnikov, F.~Petriello, and M.~Schulze, Phys. Rev. Lett. 115 (2015) 082003}
\bibitem{Cacciari:2008gp}{M.~Cacciari, G.~P.~Salam, and G.~Soyez, JHEP 04 (2008) 063}
\bibitem{Bertolini:2014bba}{D.~Bertolini, P.~C.~Harris, M.~Low, and N.~Tran, JHEP 10 (2014) 059}
\bibitem{Dasgupta:2013ihk}{M.~Dasgupta, A.~Fregoso, S.~Marzani, and G.~P.~Salam, JHEP 09 (2013) 029}
\bibitem{Larkoski:2014wba}{A.~J.~Larkoski, S.~Marzani, G.~Soyez, and J.~Thaler, JHEP 05 (2014) 146}
\bibitem{Moult:2016cvt}{I.~Moult, L.~Necib, and J.~Thaler, JHEP 12 (2016) 153}
\bibitem{Larkoski:2013eya}{A.~J.~Larkoski, G.~P.~Salam and J.~Thaler, JHEP 06 (2013) 108}
\bibitem{Dolen:2016kst}{J.~Dolen, P.~C.~Harris, S.~Marzani, S.~Rappoccio, and N.~Tran, JHEP 05 (2016) 156}
\bibitem{CMS-PAS-BTV-15-002}{CMS Collaboration, CMS-BTV-15-002 (2015), https://cds.cern.ch/record/2195743?ln=it}

 




\end{thebibliography}
\end{document}